\documentclass[a4paper]{jpconf}
\usepackage{graphicx}
\usepackage{amssymb,epsfig}
\begin{document}
\title{Partially ionized plasmas in electromagnetic fields}

\author{D Kremp$^1$, D Semkat$^1$, Th Bornath$^1$, M Bonitz$^2$, M Schlanges$^3$ and\\ P Hilse$^3$}

\address{$^1$ Universit\"at Rostock, Institut f\"ur Physik, D--18051 Rostock}
\address{$^2$ Christian--Albrechts--Universit\"at zu Kiel, Institut f\"ur Theoretische Physik und Astrophysik, D--24098 Kiel}
\address{$^3$ Ernst--Moritz--Arndt--Universit\"at Greifswald, Institut f\"ur Physik, D--17489 Greifswald}

\ead{dietrich.kremp@uni-rostock.de}

\begin{abstract}
The interaction of partially ionized plasmas with an electromagnetic field is investigated using quantum statistical methods. A general statistical expression for the current density of a plasma in an electromagnetic field is presented and considered in the high field regime. Expressions for the collisional absorption are derived and discussed. Further, partially ionized plasmas are considered. Plasma Bloch equations for the description of bound--free transitions are given and the absorption coefficient as well as rate coefficients for multiphoton ionization are derived and numerical results are presented.
\end{abstract}


\section{Introduction}

Due to the enormous progress in short-pulse laser physics, the interaction of matter with electromagnetic fields is a topic of rapidly growing interest \cite{perry94,lindl04}. In this paper we consider a partially ionized plasma under the influence of an external laser field.

A partially ionized plasma is characterized by the densities of its constituents: electrons ($n_e$), ions ($n_i$), and atoms ($n_A$). The temporal change of these densities due to ionization and recombination is determined by rate equations. The rate equation for the atom density has the following form
\begin{equation}\label{rate}
\frac{dn_{A}}{dt}=\beta n_{e}n_{e}n_{i}-\alpha n_{e}n_{A}+\beta^Rn_en_i-\alpha^Rn_A.
\end{equation} 
Here, $\alpha $ and $\beta $ are the collisional rate coefficients for the reaction $A+e\rightleftharpoons e+e+i$, and $\alpha^R $ and $\beta^R $ are the radiative rate coefficients according to $A+\hbar\omega\rightleftharpoons e+i$. In the field-free case, $\alpha^R $ and $\beta^R $ vanish. The collisional coefficients for dense plasmas, intensively discussed in \cite{KSK2005,SBK88,BS93}, are essentially determined by the interaction. The density dependence of $\alpha$ can be seen in Fig.~1 \cite{KSK2005}.
\begin{figure}[h]
\begin{minipage}[t]{5.5cm}
\epsfig{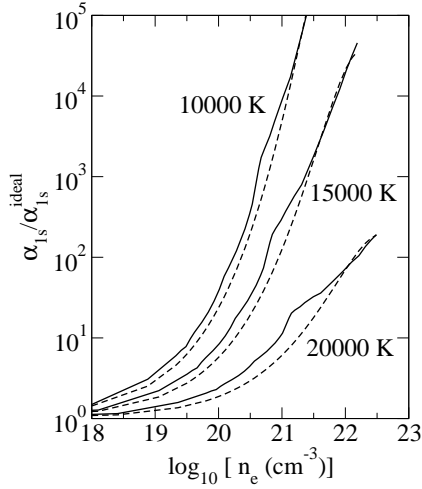}
\end{minipage}
\hfill
\begin{minipage}[b]{9.5cm}
\caption{Impact ionization coefficient for the atomic ground state in a hydrogen plasma as a function of the free electron density at several temperatures. Dashed lines: simple analytical formula for the ionization cross section proposed in \protect\cite{SBK88}, solid lines: using numerical results for the ionization cross section \protect\cite{BS93,KSK2005}.
\vspace{2.0cm}

\label{fig1}}
\end{minipage}
\end{figure}

In the stationary case, from the rate equation a mass action law follows which determines the equilibrium composition of the partially ionized plasma. This is shown in Fig.~2 \cite{KSK2005}. A characteristic feature is the strong increase of the degree of ionization ($\alpha_e$), well-known as density ionization (Mott effect).

Under the action of an external electromagnetic field given by, e.g.,
\begin{equation}  \label{monochrom}
\mathbf{A}(t)=\mathbf{A}_{0}\sin (\omega t),\;\;\mathbf{E}(t)=-%
\frac{\partial }{\partial t}\mathbf{A}(t)=\mathbf{E}_{0}\cos (\omega t),\;%
\mathbf{E}_{0}=-\omega \mathbf{A}_0,
\end{equation}
the situation is more complicated. Several additional interesting processes occur, e.g. (i) field induced transitions between bound and scattering states, i.e. radiation ionization and recombination described by $\alpha^R $ and $\beta^R $, (ii) excitation and deexcitation processes of atoms in the plasma, and (iii) processes between scattering states like (inverse) bremsstrahlung (collisional absorption).

All these processes are connected with an energy exchange between plasma and field which is determined by the electrical current density $\mathbf{j}$,
\begin{equation}
\frac{dW^{\mathrm{kin}}}{dt}+\frac{dW^{\mathrm{pot}}}{dt}=\mathbf{j}\cdot
\mathbf{E}\,,
\end{equation}
i.e., the change of the total energy of the system of particles is equal to $%
\mathbf{j}\cdot \mathbf{E}$ which is in turn the energy loss of the
electromagnetic field due to Poynting's theorem. From the quantity $\mathbf{j}$ we get, as is well known,
further important quantities like the polarization $d\mathbf{P}/dt=\mathbf{j}$, the absorption, reflection and refraction coefficients.
\begin{figure}[h]
\centerline{\epsfig{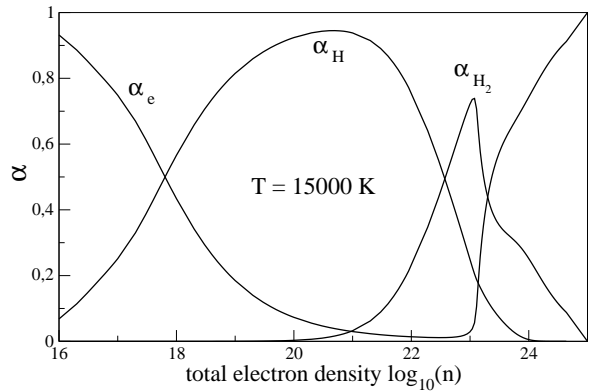}}
\caption{Equilibrium composition of a hydrogen plasma at $T=15000$ K where $\alpha_a$ denotes the fraction of the density of free particles of the species $a$ to the total electron density.\label{fig2}}
\end{figure}

A central quantity, therefore, is the electrical current density defined by
\begin{equation}
\mathbf{j}(t)=\sum\limits_{a}\ \int \frac{d{\bf p}}{(2\pi \hbar )^{3}}\mathrm{\ }%
\frac{\mathbf{p}_{a}}{m_{a}}f_{a}(p_{a},t),
\end{equation}
where $f_{a}(p_{a,}t)$ is the single-particle gauge invariant Wigner function
of the species a. For the determination of $f_{a}$ we start from the
kinetic equation \cite{kremp99,bonitz}
\begin{equation}
\left\{ \frac{\partial }{\partial t}+e_{a}\mathbf{E}(t)\cdot \nabla
_{k_{a}}\right\} {f}_{a}(\mathbf{k}_{a},t)={I}_{a}(\mathbf{k}%
_{a},t)\,,  \label{Beq}
\end{equation}%
where the collision integral is given by (${\cal V}$ being the volume and $V_{ab}$ the interaction potential)
\begin{eqnarray}
I_{a}(\mathbf{p}_a,t) 
&=&\sum_{b}\frac{1}{\mathcal{V}}\int \frac{d \mathbf{p}_b}{(2\pi \hbar )^{3}}
\left\langle
\mathbf{p}_{a}\mathbf{p}_{b}\left\vert \left[V_{ab},F_{ab}(t)\right]\right\vert \mathbf{p}_{b}%
\mathbf{p}_{a}\right\rangle\nonumber\\
&=&\sum_{b}\frac{1}{\mathcal{V}}\int \frac{d \mathbf{q}\,d\mathbf{p}_{1}%
\,d\mathbf{p}_{2}}{(2\pi \hbar )^{6}}[\delta (\mathbf{p}_a-\mathbf{p}_{1}-\mathbf{q}%
)-\delta (\mathbf{p}_a-\mathbf{p}_{1})]  \label{Collint}  \nonumber\\
&&\times V_{ab}(\mathbf{q})F_{ab}(\mathbf{p}_{1},\mathbf{p}_{1}+\mathbf{q%
},\mathbf{p}_{2},\mathbf{p}_{2}-\mathbf{q},t)\,.
\end{eqnarray}
The two-particle density matrix is connected with a two-particle correlation function by
\begin{eqnarray}
F_{ab}(t)&=&(i\hbar)^2 g^<_{ab}(12 t,1'2't)
=(i\hbar)^2 g^<_{a}(1 t,1't)\,g^<_{b}(2 t,2't) \pm \mbox{exch.}
+ (i\hbar)^2 g^{<\,{\rm corr}}_{ab}(12 t,1'2't)\nonumber
\,,
\end{eqnarray}
where we denoted $1={\bf r}_1,s_1$ etc.
For special physical situations, it is easier to derive appropriate approximations for
other functions. In place of the correlation part of $g_{ab}^<$, we can put any single-time
two-particle correlation function, especially also $L_{ab}^<-L^{0\,<}_{ab}$ which is the
correlation function of density fluctuations \cite{bornath03,springer_morawetz},
\begin{eqnarray}
(i\hbar)L^<_{ab}(11't,22't')&=&
\left\langle \delta{\hat \rho}_b(22't')\,\delta{\hat \rho}_a(11't)
\right\rangle\nonumber\\
\mbox{with}\qquad \delta{\hat \rho}_a(11't)&=&\Psi_a^+(1',t)\Psi_a(1,t)
-\langle\Psi_a^+(1',t)\Psi_a(1,t)\rangle
\end{eqnarray} 
In the momentum balance, there are contributions from collisions between different species only.
For the case $a\ne b$, we have ($a=e,i$)
\begin{eqnarray}
F_{ab}(t)&=&
(i\hbar)^2 g^<_{a}(1 t,1't)\,g^<_{b}(2 t,2't) + (i\hbar) L^{<}_{ab}(11' t,2'2't)\,.
\end{eqnarray}

Explicit expressions for the collision integral were given in the classical
case first  by Silin \cite{silin64} and later by Klimontovich \cite{klimontovich75}
and in the quantum mechanical case in the papers \cite{kremp99,bonitz99}. In generalization to the well-known Boltzmann-like collision integrals, they describe short-time processes, i.e., they are non-Markovian \cite{SKB99,SKB02,SBK03,KSB05}.

In a partially ionized plasma, we have to account for bound and scattering
states and transitions between them caused by the external
field. Therefore, it is useful to expand $F_{ab}$ with respect to
the eigenstates of the unperturbed two-particle Hamiltonian (the spin dependence is not written explicitly)
\begin{equation}
F_{ab}(\mathbf{p}_{1},\mathbf{p}_{1}+\mathbf{q},\mathbf{p}_{2},\mathbf{p}_{2}-\mathbf{q},t)
=\sum_{\mathbf{P}\alpha,\mathbf{P}^{\prime} \alpha^{\prime}}
\left\langle \mathbf{p}_{1}\mathbf{p}_{2}\right\vert \alpha \mathbf{P}\rangle \left\langle
\mathbf{P}\alpha \left\vert F_{ab}(t)\right\vert \alpha ^{\prime
}\mathbf{P}^{\prime }\right\rangle \left\langle \mathbf{P}^{\prime }\alpha ^{\prime
}\right\vert \mathbf{p}_{1}+\mathbf{q},\mathbf{p}_{2}-\mathbf{q}\rangle  \label{exp}
\end{equation}
with $\alpha =\{n,l,m\}$ for bound states and $\alpha =\mathbf{p}$ for scattering states. ${\bf P}$ is the center-of-mass momentum.
As we will show lateron, due to the external field, there are nonvanishing nondiagonal elements $%
F_{ab}^{\alpha \alpha ^{\prime }}$.  They are connected with transitions
between different states $\alpha$ and $\alpha ^{\prime}$.

The balance equation for the
electrical current density follows in well--known manner from Eq.~(\ref{Beq}) \cite{bornath03}
\begin{eqnarray}
\frac{d}{dt}\mathbf{j}_{a}(t)-n_{a}\frac{e_{a}^{2}}{m_{a}}%
\,\mathbf{E}(t)&=&\frac{1%
}{2i\hbar}\sum_{b}\int \frac{d{\bf p}_{a}d{\bf p}_{b}}{(2\pi \hbar )^{6}}\left(\frac{e_{a}%
\mathbf{p}_{a}}{m_{a}}+\frac{e_{b}\mathbf{p}_{b}}{m_{b}}\right)\ \left\langle
\mathbf{p}_{a}\mathbf{p}_{b}\left\vert \left[V_{ab},F_{ab}(t)\right]\right\vert \mathbf{p}_{b}%
\mathbf{p}_{a}\right\rangle\nonumber\\
&=&
\sum_{b\neq a}\int \frac{d\mathbf{q}}{(2\pi \hbar )^{3}}\,\frac{%
e_{a}\mathbf{q}}{m_{a}}\,V_{ab}(q)\,L_{ba}^{<}(\mathbf{q};t,t)\,.
\label{Currbal}
\end{eqnarray}
Further progress is connected with the determination of the functions $L_{ba}^{<}$ or $F_{ab}$, respectively.

In the following, we will consider the free--free transition using approximations for $L_{ba}^{<}$ in the next section. Then
the current is calculated from Eq.~(\ref{Currbal}), and the collisional absorption (inverse
bremsstrahlung) in the linear response regime as well as in the high--field case is
determined and discussed.

In the third section, bound states are included. The equation of motion for $g_{ab}^<$ is the
Bethe--Salpeter equation (BSE) for plasmas which has been discussed in detail in \cite{BKS99}.
We will derive plasma Bloch equations which form a system of equations for the determination
of the matrix elements of (\ref{exp}). Here, the equations for the diagonal elements describe the
time evolution of the occupation of bound and scattering states, respectively, driven by the non-diagonal elements. The coupled system, therefore, describes the ionization kinetics of a partially ionized plasma.

\section{Collisional absorption in a fully ionized plasma}

Let us first consider the energy exchange between field and plasma due to transitions between scattering states. This problem has been dealt with in several papers, first for classical plasmas, e.g. \cite{oberman62,silin64,klimontovich75,decker94,mulser}. Quantum mechanical treatments were given in more recent papers, for the strong field case, e.g., \cite{kull01,bshk01,hazak}, and in linear response theory \cite{reinholz00}.

Starting point for our investigation is Eq.~(\ref{Currbal}). The r.h.s. of this equation is determined by the function $L^<$. This quantity follows from the Bethe--Salpeter equation on the Keldysh contour
\begin{equation}
L_{ab}(12,1^{^{\prime }}2^{^{\prime }})=\Pi _{ab}(12,1^{^{\prime
}}2^{^{\prime }})+\sum_{cd}\int\limits_{C}d3d4\Pi
_{ac}(13,1^{^{\prime }}3^{+})V_{cd}(34)L_{db}(42,4^{+}2^{^{\prime
}})\,.
\end{equation}
In a plasma in a strong laser field, an approximation in lowest order of $V_{ie}$ is appropriate because the coupling between species
with different charges is weaker than their coupling to the field. The
coupling between particles with equal charges in the subsystem, however, is not
affected by the field.
Then the so-called
polarization functions $\Pi_{ab}$ can adopted to be approximately diagonal,
$\Pi_{ab}=\delta_{ab}\Pi_a$. The solution has now the following structure (for brevity all
arguments are suppressed)
\begin{eqnarray}
L_{ei}=\mathcal{L}_{ee}\ V_{ei}\ \mathcal{L}_{ii}\ ;\ \ \ \ \ \ \mathcal{L}%
_{aa}=\Pi _{a}+\Pi _{a}\,V_{aa}\,\mathcal{L}_{aa};\ \ \ \ (a=e,i).
\end{eqnarray}
Going back from the contour to the physical time axis using the Langreth--Wilkins rules we find for the correlation functions of the density fluctuation \cite{springer_morawetz}
\begin{eqnarray}\label{appro}
 L^\gtrless_{ei}({\bf q}; t,t')=\int d{\bar t}\, \Big[
{\cal L}_{ee}^\gtrless({\bf q}; t,{\bar t})\,V_{ei}(q)
\,{\cal L}_{ii}^A({\bf q}; {\bar t},t')
+ {\cal L}_{ee}^R({\bf q}; t,{\bar t})\,V_{ei}(q)\,
{\cal L}_{ii}^\gtrless({\bf q};{\bar t},t')\Big]\,.
\end{eqnarray}
For the electron current (\ref{Currbal}) there follows that
\begin{eqnarray}\label{curr}
  \frac{{\rm d}}{{\rm d} t} {\bf j}_e(t) - n_e
\frac{e_e^2}{m_e}\,{\bf E}(t)
&=&{\rm Re}\int \frac{d^3 q}{(2\pi\hbar)^3}
\,\frac{e_e \bf q}{m_e \hbar}\,V_{ei}(q)\,2\pi i\,
\int_{t_0}^t d{\bar t} \Big[\,{\cal S}_{ee}({\bf q};t,{\bar t})V_{ei}(q) {\cal L}_{ii}^A({\bf q}; {\bar t},t)
\nonumber\\
&&
+{\cal L}_{ee}^R({\bf q}; t,{\bar t})\,V_{ei}(q)\,
{\cal S}_{ii}({\bf q};{\bar t},t)
\Big]\,,
\end{eqnarray}
where we introduced the dynamical structure factor
\begin{eqnarray}
2\pi {\cal S}_{aa}({\bf q};t,{\bar t})=\frac{i\hbar}{2}\Big[
{\cal L}^>_{aa}\,({\bf q}; t,{\bar t})+
{\cal L}^<_{aa}\,({\bf q}; t,{\bar t})\Big]\,.
\end{eqnarray}
The functions in the collision term depend on the current and the electrical field, respectively.
This dependence can be made explicit if one assumes that each subsystem (electrons and ions) is in
local equilibrium with a temperature $T_a$ with respect to
a coordinate frame moving with the mean velocity ${\bf u}_a(t)$ \cite{KSK2005,bornath04}.
The transformation between such a coordinate system and a system at
rest is given by
$\tilde{\bf r}={\bf r}-\int_{t_0}^t {\rm d}{\bar t}\,{\bf u}_a({\bar t})$.
The Fourier transforms in the two coordinate systems are connected by
\begin{eqnarray}
{\cal L}_{aa}({\bf q},t_1 t_2)
&=&e^{-\frac{i}{\hbar}{\bf q}\cdot
\int_{t_2}^{t_1} {\rm d}{\bar t}\,{\bf u}_a({\bar
t})}\,
\tilde{\cal L}_{aa}({\bf q}, t_1-t_2)\,,
\end{eqnarray}
where $\tilde{\cal L}_{aa}$ denotes the local equilibrium function depending on the time difference only.
One gets (omitting the tilde from now on)
\begin{eqnarray}\label{bal1}
\frac{{\rm d} }{{\rm d}t}{\bf j}(t)
&=&\varepsilon_0 \omega^2_p\,{\bf E}(t)-{\rm Re}\int \frac{d^3 q}{(2\pi\hbar)^3}\,{\bf q}
\,\frac{e_e}{m_e}\,V_{ie}(q)\,\frac{2\pi}{i\hbar}
\int_{t_0}^t d{\bar t}\, \Big[\,{\cal S}_{ee}({\bf q};t-{\bar
t}) V_{ei}(q)
\\
&&\times {\cal L}_{ii}^A({\bf q}; {\bar t}-t)+
{\cal L}_{ee}^R({\bf q}; t-{\bar t})
\,V_{ei}(q)\,{\cal S}_{ii}({\bf q};{\bar t}-t)
\Big]\exp\left\{-\frac{i}{\hbar}\frac{1}{n_ee_e}
{\bf q}\cdot\int_{\bar t}^t d{\bar t}_1\,
{\bf j}({\bar t}_1)\right\}
\nonumber\,.
\end{eqnarray}
Thus the source term in the current balance equation depends on the current itself in a non-linear
way.

We will consider this non-linear equation in the following in two limiting cases.

\subsection{Linear response case}

Let us first consider the energy exchange in a weak laser field. To describe the situation we introduce the quiver velocity
\begin{equation}
\mathbf{v}_0=\frac{e_e\mathrm{E}_0}{m_e\omega}.
\end{equation} 
Then the weak field case is characterized by $v_0/v_{\rm th}\ll 1$ where $v_{\rm th}$ is the electron thermal velocity. Therefore, the situation corresponds to the linear response regime. In this case we can expand the exponential
function in Eq.~(\ref{bal1}),
\begin{eqnarray}
\exp \left\{ -\frac{i}{\hbar }\frac{1}{n_{e}e_{e}}\mathbf{q}\cdot \int_{\bar{%
t}}^{t}d{\bar{t}}_{1}\,\mathbf{j}({\bar{t}}_{1})\right\} \approx 1--\frac{i}{%
\hbar }\frac{1}{n_{e}e_{e}}\mathbf{q}\cdot \int_{\bar{t}}^{t}d{\bar{t}}_{1}\,%
\mathbf{j}({\bar{t}}_{1}).
\end{eqnarray}
Adopting a harmonic time dependence of the electric field, we  have for the current
${\bf j}(t)={\bf j}(\omega)e^{-i\omega t}+{\bf j}^*(\omega)e^{i\omega t}$ and get from the above
equation
\begin{eqnarray}
{\bf j}(\omega)=\frac{\varepsilon_0 \omega^2_p}{-i\omega + \nu_{ei}(\omega)}{\bf E}(\omega),
\end{eqnarray}
which is a generalized Drude equation with a complex electron-ion collision frequency $\nu_{ei}$ given by
\begin{eqnarray}\label{nyeilr}
\nu_{ei}(\omega)
= i\frac{n_i e_i^2}{6\pi^2n_e m_e \hbar^3}\frac{1}{\omega} \int\limits_0^{\infty}d q
\,q^4\,{\cal S}_{ii}(q) V(q) \Big[\varepsilon^{-1}_{ee}({\bf q};\omega)
- \varepsilon^{-1}_{ee}({\bf q};0)\Big]\,
\end{eqnarray}
with the dielectric function of the electron subsystem
\begin{eqnarray}
\varepsilon^{-1}_{ee}({\bf q};\omega)
&=&1+
e^2 V(q)\,{\cal L}_{ee}({\bf q};\omega)\,,
\end{eqnarray}
the static ion-ion structure factor ${\cal S}_{ii}(q)$, . Using for $\varepsilon^{-1}_{ee}$ the RPA,
this is the well-known expression first given by
Bekefi \cite{bekefi66} and later for instance by Reinholz et al. \cite{reinholz00}. One gets immediately the
dynamical conductivity
\begin{eqnarray}
\sigma(\omega)=\frac{\varepsilon_0 \omega^2_p}{-i\omega + \nu_{ei}(\omega)}.
\end{eqnarray}
\begin{figure}[h]
\begin{minipage}[t]{8cm}
\epsfig{figure=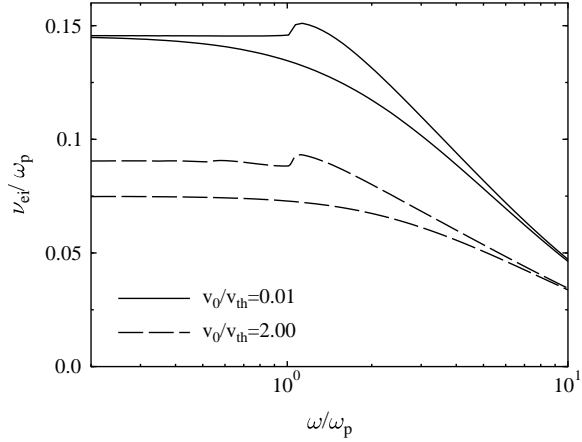,clip=true,width=8cm}
\end{minipage}
\hfill
\begin{minipage}[b]{8cm}
\caption{Electron--ion collision frequency as a function of the laser
frequency for a hydrogen plasma ($n_e= 10^{21}{\rm cm}^{-3}$;
$T=10^5$ K) for two different field strengths. The upper curve of
each pair corresponds to the full dynamical screening, the lower one
to the static screening approximation. \label{omi}}
\vspace{1.5cm}

\end{minipage}
\end{figure}
In the limit of high frequencies, one gets
$\sigma(\omega)\approx (\varepsilon_0 \omega^2_p/\omega^2)[i\omega +\nu_{ei}(\omega)]$ \cite{oberman62},
and the averaged absorbed energy is given by
\begin{eqnarray}\label{nuei}
\langle {\bf j}\cdot {\bf E} \rangle\equiv \frac{1}{T}\int_{t-T}^{T} dt'  {\bf j}(t')\cdot {\bf E}(t')
=\frac{\omega^2_p}{\omega^2} {\rm Re}\,\nu_{ei}(\omega)\,\frac{\varepsilon_0{\bf E}_0^2}{2}=
\frac{\omega^2_p}{\omega^2} {\rm Re}\,\nu_{ei}(\omega)\,\langle \varepsilon_0{\bf E}^2 \rangle \,.
\end{eqnarray}
The dependence of the quantity ${\rm Re}\,\nu_{ei}$ on the laser frequency is shown in Fig.~\ref{omi} for two different relations $v_0/v_{\rm th}$. Only the upper set of curves correspond to the linear response formula (\ref{nyeilr}). Obviously, statical and dynamical screening exhibit a qualitatively different behavior. The dynamical screening leads to a maximum in the vicinity of the plasma frequency,  $\omega\approx\omega_p$. The lower set of curves is beyond the linear response regime. This strong field case will be considered in the next paragraph.

\subsection{Strong high-frequency fields}

The strong field case is characterized by $v_0/v_{\rm th}\gtrsim 1$.
For the first time such a situation was discussed by Silin \cite{silin64} in the framework of classical kinetic theory.

For a strong high-frequency laser field a linearization with respect to the field is not possible. Instead, we follow the idea of Silin and assume that the influence of the collisions may be considered as a small perturbation compared to the external field. We decompose the current according to $\mathbf{j}=\mathbf{j}_0+\mathbf{j}_1$ and assume
\begin{eqnarray}
{\bf j}\approx {\bf j}^0=\sum_a \frac{e_a^2n_a}{m_a}\int_{t_0}^t {\rm d}t' {\bf E}(t'\,).
\end{eqnarray}
For a harmonic electric field, ${\bf E}={\bf E}_0 \cos \omega t$,
the exponential factor in Eq.~(\ref{bal1}) can be expanded into a Fourier series. The current
balance is given then by
\begin{eqnarray}\label{jbil}
\hspace*{-10ex}&&\frac{\rm d}{\rm d t} {\bf j}_e(t) -n_e\frac{e_e^2}{m_e}\,{\bf E}(t)
 = {\rm Re}\,\int\frac{d^3 q}{(2\pi\hbar)^3}\,\frac{2\pi e_e{\bf q}}{m_e \hbar}\,
V^2_{ei}(q)\,
\sum_{m} \sum_n (-i)^{m+1} J_n\left(\frac{{\bf q}\cdot{\bf v}_{0}}{\hbar
\omega} \right)
J_{n-m}\left(\frac{{\bf q}\cdot{\bf v}_{0}}{\hbar
\omega} \right)
\nonumber\\
\hspace*{-10ex}&&\qquad\times e^{i m\omega t}\,\,
\int\limits_{-\infty}^{\infty}
\frac{d{\bar \omega}}{2\pi}\,
\left[{\cal S}_{ee}({\bf q};{\bar \omega}-n\omega)\,
{\cal L}_{ii}^A({\bf q};{\bar \omega})
+{\cal L}_{ee}^R({\bf q}; {\bar \omega}-n\omega)
\,{\cal S}_{ii}({\bf q};{\bar \omega})
\right]
\end{eqnarray}
with the one-component structure factors and response functions ${\cal S}_{aa}$ and
${\cal L}_{aa}$, respectively \cite{bornath03,springer_morawetz}. We will assume in the following of this
section that the subsystems are in local thermodynamic equilibrium with temperatures
$T_e$ and $T_i$, respectively (the influence of non-Maxwellian distribution functions was considered in \cite{dresden}, and a numerical solution of a kinetic equation for a strong laser field was performed in \cite{haberland}).
$J_l$ is the Bessel function of $l$th order. In the above equation, electron and ion functions contribute
formally equal to the screening.
The ion functions, however, are localized in the low-frequency region,
i.e., for a high-frequency electric field, ${\bar \omega}$ can be neglected in
comparison with $n\omega$. In this case, the first term in the
brackets in Eq. (\ref{jbil}) vanishes because $\int d{\bar \omega} {\cal L}_{ii}^A({\bf q};{\bar
\omega})=0$, and for the current it follows that
\begin{eqnarray}\label{harm}
{\bf j}_e(t)-\int^t_{-\infty}d{\bar t}\, \frac{n_ee_e^2}{m_e}{\bf
E}({\bar t})
&=& {\rm Re}\,
\,\int\frac{d^3 q}{(2\pi\hbar)^3}
\,\sum_{m}\sum_n \frac{e_e}{m_e}\frac{{\bf q}}{m\hbar\omega}\,V^2_{ei}(q)\,
(-i)^{m+2}e^{i m\omega t}
\nonumber\\
&&
\times
J_n\left(\frac{{\bf q}\cdot{\bf v}_0}{\hbar
\omega} \right)
J_{n-m}\left(\frac{{\bf q}\cdot{\bf v}_0}{\hbar
\omega} \right)
{\cal L}_{ee}^R({\bf q};-n\omega)
\,n_i\,{\cal S}_{ii}({\bf q})
\,
\end{eqnarray}
where screening by the ions is accounted for by the static structure
factor ${\cal S}_{ii}({\bf q})$ defined by
\begin{eqnarray}
{\cal S}_{ii}({\bf q})&\equiv&
\frac{1}{n_i}\int d{\bar \omega}\, {\cal S}_{ii}({\bf q},{\bar \omega})
=1 + n_i \int d^3r\, [g_{ii}({\bf r})-1]\,
e^{-\frac{i}{\hbar}{\bf q}\cdot{\bf r}}\,,
\end{eqnarray}
where $g_{ii}$ is the pair distribution function.
${\cal L}_{ee}^R$ is the exact density response function of the
electron subsystem. Appropriate approximations can be
expressed via local field corrections \cite{springer_morawetz}.

Equation (\ref{harm}) is clearly an expansion of the current in terms of higher harmonics of the laser frequency
\begin{equation}
{\bf j}_e(t)-\int^t_{-\infty}d{\bar t}\, \frac{n_ee_e^2}{m_e}{\bf E}({\bar t})
= \sum\limits_{m=-\infty}^{\infty} {\bf j}_m(t) {\rm e}^{im\omega t}.
\end{equation} 
The Fourier coefficients ${\bf j}_m$ of the current (with ${\bf j}_m={\bf j}^*_{-m}$) can be easily identified from (\ref{harm}). One can show that only the odd harmonics are allowed due to the symmetry of the interaction, cf.~\cite{bshk01}. The appearance of higher harmonics in strong laser fields was first observed by Silin \cite{silin64} and is a very interesting effect.

Let us now consider again the cycle averaged energy (\ref{nuei}). This quantity is now given by
\begin{eqnarray}\label{jE1}
\langle {\bf j}\cdot {\bf E} \rangle\equiv \frac{1}{T}\int_{t-T}^{T} dt'  {\bf j}(t')\cdot {\bf E}(t')
= {\bf E}_0 \cdot {\rm Re}\,{\bf j}_1
=\frac{\omega^2_p}{\omega^2} {\rm Re}\,\nu_{ei}(\omega)\,\frac{\varepsilon_0{\bf E}_0^2}{2} \,.
\end{eqnarray}
Interesting is here the restriction of the sum over $m$ to its first term, i.e., higher harmonics do not contribute to the average energy exchange.

Using the explicit expression for ${\rm Re}\,{\bf j}_1$ we obtain
\begin{eqnarray}\label{jE}
\left\langle{\bf j}\cdot{\bf E}\right\rangle \,
&=& n_i\int\frac{d^3 q}{(2\pi\hbar)^3} \,V_{ii}(q)\,
{\cal S}_{ii}({\bf q}, T_i)
\sum_{n=-\infty}^{\infty}  \, n\omega\,
J^2_n\left({\textstyle\frac{{\bf q}\cdot{\bf v}_0}{\hbar \omega}}
\right) {\rm Im}\,{\varepsilon}_{ee}^{-1}({\bf q},-n\omega,T_e).
\end{eqnarray}
Finally, the effective electron--ion collision frequency can be obtained using Eq.~(\ref{jE1}).

In Eq.~(\ref{jE}), ${\varepsilon}_{ee}^{-1}$ is the full quantum mechanical dielectric function of the electrons, e.g. the RPA one.
The physical meaning of the sum over $n$ follows from the frequency argument $n\omega$ of the dielectric function. It indicates that every term of the sum describes the absorption of $n$ photons (multi-photon absorption).

In order to compare with the classical theory, it is advantageous to consider the non-degenerate case in which some integrations can be done analytically. For the case of a Maxwellian distribution function, we get
\begin{eqnarray}\label{jEmaxwell}
\langle \,\mathbf{j}\cdot \mathbf{E}\,\rangle  &=&\frac{8\sqrt{2\pi }%
Z^{2}e^{4}n_{e}n_{i}\sqrt{m_{e}}}{(4\pi \varepsilon _{0})^{2}(k_{B}T)^{3/2}}%
\,\omega ^{2}\,\sum_{n=1}^{\infty }\,n^{2}\,\int\limits_{0}^{\infty }\frac{dk%
}{k^{3}}\,\frac{1}{|\varepsilon (k,n\omega )|^{2}}\, \nonumber\\
&&\times e^{-\frac{n^{2}m_{e}\omega
^{2}}{2k_{B}T\,k^{2}}}\,e^{-\frac{\hbar
^{2}k^{2}}{8m_{e}k_{B}T}}\,\,\frac{\sinh {\frac{n\hbar \omega }{2k_{B}T}}}{%
\frac{n\hbar \omega
}{2k_{B}T}}\,\,\int\limits_{0}^{1}dz\,J_{n}^{2}\left(
\frac{eE_{0}k}{m_{e}\omega ^{2}}\,z\right) \,.
\end{eqnarray}
Quantum effects, marked by $\hbar$, occur here in several places. The first place is one of the exponential functions describing the quantum diffraction effects at large momenta $k$. This exponential function ensures the convergence of the integral. The second place is the sinh term which is connected with the Bose statistics of multiple photon emission and absorption. Finally, quantum effects enter via the calculation of dielectric function itself.

In the limit $\hbar\to 0$ we get from Eq.~(\ref{jE}) well-known classical results derived for the first time by Klimontovich \cite{klimontovich75} and later by Decker et al. \cite{decker94}. The classical formulas exhibit the well-known problem of divergencies at large $k$ which has to be overcome by some cut-off procedure. In contrast, in our quantum approach, no divergencies exist.

\begin{figure}[bth]
\begin{minipage}[t]{8cm}
\epsfig{figure=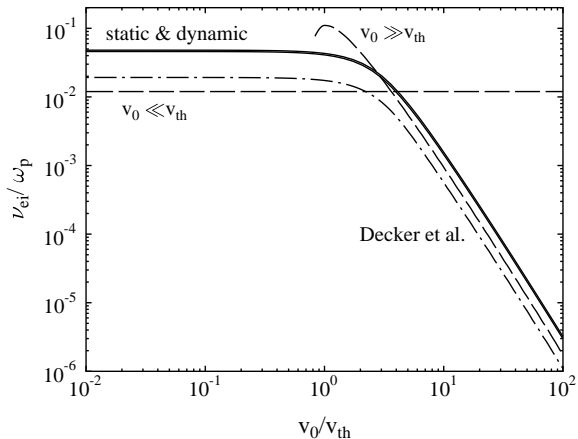,clip=true,width=8cm}
\end{minipage}
\hfill
\begin{minipage}[b]{8cm}
\caption{Real part of the electron--ion collision frequency as a
function of the quiver velocity $v_0=eE/\omega m_e$ for a hydrogen plasma in a laser field ($Z=1;
n_e=10^{22}$ cm$^{-3}; T=3 \cdot10^{5}$ K; $\omega/\omega_p=5$). For
comparison, results of Decker et al. (dash-dotted line) and the
asymptotic formulae of Silin \cite{silin64} (dashed line) are given. \label{b4a}}
\vspace{1.5cm}

\end{minipage}
\end{figure}

\begin{figure}[bth]
\begin{minipage}[t]{8cm}
\epsfig{figure=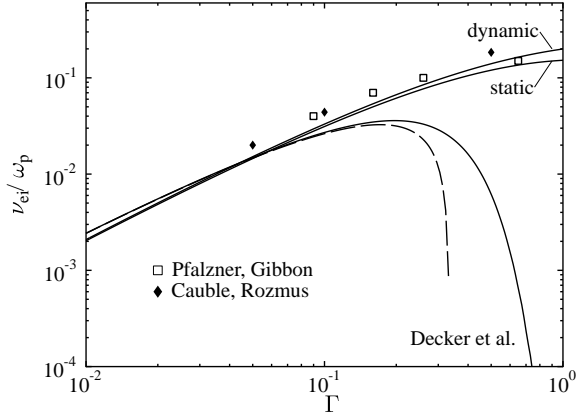,clip=true,width=8cm}
\end{minipage}
\hfill
\begin{minipage}[b]{8cm}
\caption{ Electron--ion collision frequency as a function of the
coupling parameter $\Gamma$ for a hydrogen plasma in a laser field
($Z=1; v_0/v_{th}=0.2; n_e= 10^{22}$ cm$^{-3}; \omega/\omega_p=3$).
A comparison is given with the theory of Decker et al. and with the
asymptotic formula of Silin \cite{silin64} (dashed line) is given. Furthermore, results of Cauble and Rozmus \protect\cite{cauble85} and simulation data of Pfalzner and Gibbon \protect\cite{pfalzner98} are shown. }
\label{nue_gam02}
\vspace{1.5cm}

\end{minipage}
\end{figure}

Unfortunately, a further analytical simplification is only possible in limiting cases. Therefore, we present in the following some results of the numerical evaluation of Eq.~(\ref{jE}) \cite{SBKH03,bornath05}.
In Fig.~\ref{b4a}, the collision frequency is shown as a function of the quiver velocity. For comparison, results from Silin \cite{silin64} and Decker \cite{decker94} are given. Our quantum expression (\ref{jE}) was evaluated with the full dynamical and the static RPA dielectric function.
The frequency dependence has been shown already in Fig.~\ref{omi}. Again, we observe the resonance maximum at the plasma frequency.
The collision frequency as a function of the coupling parameter
$\Gamma=\left(\frac{4\pi n}{3}\right)^{1/3}\,\frac{e^2}{k_{\rm B}T}$ is drawn in Fig.~\ref{nue_gam02}. According to this figure our quantum mechanical results are in good agreement with numerical simulations in this special situation. In contrast, the classical theories break down for $\Gamma\gtrsim 0.2$. This behavior results from the cut-off procedure at large momentum $k$ inherent in classical approaches.

\begin{figure}[bth]
\begin{minipage}[t]{8cm}
\epsfig{figure=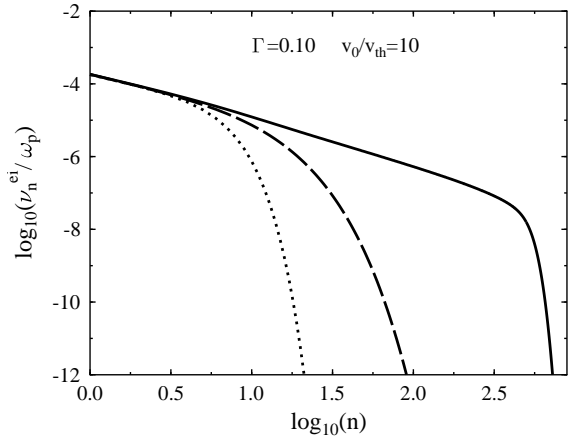,clip=true,width=8cm}
\end{minipage}
\hfill
\begin{minipage}[b]{8cm}
\caption{The $n-$photon contributions to  $\nu_{ei}$ vs. photon number $n$ in a hydrogen
plasma ($n_e= 10^{22}{\rm cm}^{-3}$; $\omega/\omega_p=5$;
$\Gamma=0.1$) for $v_0/v_{th}=10$. Present approach (solid line),
$\sinh$ term neglected (dashed line); classical dielectric theory
(dotted line) \label{x10g01}}
\vspace{1.5cm}

\end{minipage}
\end{figure}

The influence of the Bose character of the photons [hyperbolic sine term in (\ref{jEmaxwell})] can be seen in Fig.~\ref{x10g01}. An interesting consequence is the plateau-like behavior of the collision frequency with a subsequent sharp drop down in dependence of the photon number. Neglecting the sinh term destroys this behavior.


\section{Bound--free transitions. Ionization kinetics}
\subsection{Basic equations}\label{basic}

So far we have considered transitions between scattering states. Now we take into account bound states and consider bound--free transitions in a partially ionized plasma, i.e., radiation ionization and recombination processes. Therefore, our task is to develop a kinetic description of these processes, especially we have to determine rate and absorption coefficients of radiation ionization. To
investigate the problem of build-up
and decay of bound states, an appropriate starting point of our
considerations is the equation of motion of the two-particle correlation
function $g_{ab}^{<}$, well-known as Bethe--Salpeter equation. In order to take into account bound
states, we have to consider the two-particle correlation function $g_{ab}^{<}$ at least in binary
collision approximation \cite{SKB02,BSKF03}. Then the BSE can be written in the form \cite{BKS99,KSK2005}
\begin{equation}
\left[ i\hbar \frac{\partial }{\partial t}-{H}
_{ab}^{0}-U_{ab}-N_{ab}V_{ab}-\left( \Sigma _{a}^{\mathrm{HF}}(t)+\Sigma _{b}^{%
\mathrm{HF}}(t)\right) \right] g_{ab}^<(t,t^{\prime })=0.  \label{BSE2}
\end{equation}%
Here, $\Sigma _{a}^{\mathrm{HF}}(t)$ is the Hartree--Fock self-energy, and $N_{ab}=1-F_a-F_b$ the Pauli blocking factor.
It is convenient to introduce relative
and center-of-mass coordinates $\mathbf{p}$ and $\mathbf{P}$, respectively, which yields
\begin{equation}\label{hamilt}
H_{ab}^{0}=\frac{\mathbf{P}^{2}}{2M}+\frac{\mathbf{p}^{2}}{2\mu };\ \ \
U_{ab}=-\mathbf{A}\cdot \left( \frac{e_{t}\mathbf{P}}{M}+\frac{%
e_{r}\mathbf{p}}{\mu }\right) +\frac{A^{2}}{2}\left( \frac{e_{t}^{2}}{M}%
+\frac{e_{r}^{2}}{\mu }\right) ,
\end{equation}%
where $M$ and $\mu $ are the total and reduced masses, and $e_{t}=e_{a}+e_{b}$ the total and
$e_r$ the reduced charge defined by $Me_{r}=m_{b}e_{a}-m_{a}e_{b}$, respectively.

Now we take the time-diagonal limit of the two-particle correlation function, $F_{ab}(t)=(i\hbar )^{2} g_{ab}^{<}(t,t)$. From Eq.~(\ref{BSE2}) there follows that \cite{bonitz}
\begin{eqnarray}\label{FabL}
&&i\hbar \frac{\partial }{\partial t}F_{ab}(t)-\left[
H_{ab}^{0}+V_{ab},F_{ab}(t)\right] -\left[ U_{ab}(t),F_{ab}(t)\right]  \nonumber
\\[0.15cm]
&&-\left[ \left( F_{a}(t)+F_{b}(t)\right) V_{ab},F_{ab}(t)\right] -\left[
\left( \Sigma _{a}^{\mathrm{HF}}(t)+\Sigma _{b}^{\mathrm{HF}}(t)\right)
,F_{ab}(t)\right] =0,  \label{BSEtd}
\end{eqnarray}
which describes the kinetics of the two-particle properties like bound and
scattering states in the plasma. In Eq.~(\ref{FabL}), the first two terms represent the usual binary collision approximation, the third term contains the direct external field contribution, whereas the further terms arise from many-particle effects: Pauli blocking and Hartree--Fock (mean field + exchange) self-energy.

For the further investigation, we choose, for the two-particle density matrix, the basis representation (\ref{exp}) in terms of the bound and scattering eigenstates of the
field-free two-particle Hamiltonian defined by
\begin{eqnarray}
\left( H_{ab}^{0}+V_{ab}\right) \left\vert \mathbf{P}\alpha \right\rangle
&=&E_{\alpha \mathbf{P}}\left\vert \mathbf{P}\alpha \right\rangle ,
\label{represent} \\
\left\vert \mathbf{P}\alpha t\right\rangle &=&e^{-\frac{i}{\hbar }E_{\alpha
\mathbf{P}}t}\left\vert \mathbf{P}\alpha \right\rangle .
\end{eqnarray}%
Here, $\alpha $ denotes a complete set of quantum numbers $n,l,m\ $in the case of
bound states and the relative momentum $\mathbf{p}$ in the case of
scattering states, respectively,
\begin{equation*}
\alpha =\left\{
\begin{array}{r@{\quad}l}
\{n,l,m\} & \mbox{for bound states} \\
\mathbf{p} & \mbox{for scattering states.}%
\end{array}%
\right.
\end{equation*}

In the representation (\ref{represent}), we get for the equation (\ref{FabL}) \cite{KSB05a}
\begin{eqnarray}
\left[ i\hbar \frac{\partial }{\partial t}-\left( E_{\alpha \mathbf{P}%
}-E_{\alpha ^{\prime }\mathbf{P}}\right) \right] F_{ab}^{\alpha \alpha
^{\prime }}(\mathbf{P};t)=  
\sum\limits_{\bar{\alpha}}\left[ \hbar \Omega _{\mathrm{R}}^{\alpha {\bar{%
\alpha}}}(\mathbf{P};t)F_{ab}^{{\bar{\alpha}}\alpha ^{\prime }}(\mathbf{P}%
;t)-F_{ab}^{\alpha {\bar{\alpha}}}(\mathbf{P};t)\hbar {\tilde{\Omega}}_{%
\mathrm{R}}^{{\bar{\alpha}}\alpha ^{\prime }}(\mathbf{P};t)\right]
,  \label{BSEat}
\end{eqnarray}%
with the matrix elements given by
\begin{eqnarray}
\left\langle \alpha \mathbf{P}\right\vert F_{ab}(t)\left\vert \mathbf{P}%
^{\prime }\alpha ^{\prime }\right\rangle =F_{ab}^{\alpha \alpha ^{\prime }}(%
\mathbf{P};t)(2\pi \hbar )^{3}\delta (\mathbf{P}-\mathbf{P}^{\prime }),
\label{matelem1} \\[0.2cm]
\left\langle \alpha \mathbf{P}\right\vert U_{ab}(t)+\Sigma _{a}^{\mathrm{HF}%
}(t)+\Sigma _{b}^{\mathrm{HF}}(t)\pm \left[ F_{a}(t)+F_{b}(t)\right]
V_{ab}\left\vert \mathbf{P}^{\prime }\alpha ^{\prime }\right\rangle
\label{matelem2} \nonumber \\
=\hbar \Omega _{\mathrm{R}}^{\alpha \alpha ^{\prime }}(\mathbf{P};t)(2\pi
\hbar )^{3}\delta (\mathbf{P}-\mathbf{P}^{\prime }),  \\[0.2cm]
\left\langle \alpha \mathbf{P}\right\vert U_{ab}(t)+\Sigma _{a}^{\mathrm{HF}%
}(t)+\Sigma _{b}^{\mathrm{HF}}(t)\pm V_{ab}\left[ F_{a}(t)+F_{b}(t)\right]
\left\vert \mathbf{P}^{\prime }\alpha ^{\prime }\right\rangle
\label{matelem2a} \nonumber \\
=\hbar {\tilde{\Omega}}_{\mathrm{R}}^{\alpha \alpha ^{\prime }}(\mathbf{P}%
;t)(2\pi \hbar )^{3}\delta (\mathbf{P}-\mathbf{P}^{\prime }).
\end{eqnarray}
Here, $\hbar \Omega _{\mathrm{R}}$ is an abbreviation which can be
interpreted as a generalized Rabi energy, i.e., the field contribution
renormalized by Hartree--Fock and Pauli blocking terms. Notice that $U_{ab}$
in Eqs.~(\ref{matelem2},\ref{matelem2a}) does not contain the term quadratic
in $A$ (cf.~Eq.~(\ref{hamilt})), because it cancels out in the homogeneous
case in the commutator. The quantities defined in (\ref{matelem2},\ref{matelem2a}) are related to each other by $\left[ \hbar \Omega _{\mathrm{R}%
}^{\alpha \alpha ^{\prime }}(\mathbf{P};t)\right] ^{\ast }=\hbar {\tilde{%
\Omega}}_{\mathrm{R}}^{\alpha ^{\prime }\alpha }(\mathbf{P};t)$.

Thus, the representation chosen above leads to a transformation of (\ref{FabL}) into a matrix equation. Due to the inclusion of the scattering continuum, $F_{ab}$ is, in principle, a matrix of infinite rank. If there are $N-1$ bound states, we obtain $N^2$ coupled equations. Even if one makes use of the symmetry relation 
$\left[F_{ab}^{\alpha\alpha^{\prime}}(\mathbf{P};t)\right]^*=
F_{ab}^{\alpha^{\prime}\alpha}(\mathbf{P};t)$, there remain $\frac{N}{2}(N+1)$ equations.

Obviously, the diagonal elements $F_{ab}^{\alpha \alpha }$ are related to
the occupation numbers of the respective state. Due to the external
field we have nonvanishing nondiagonal elements $F_{ab}^{\alpha \alpha
^{\prime }}$. They are connected with transitions between two states $%
\alpha \ $and $\alpha ^{\prime }$. Here, we have to consider several
situations:

(i) If both $\alpha $ and $\alpha ^{\prime }$ denote bound states, the
equations describe excitation and deexcitation processes of atoms in the
plasma. Then we recover the familiar atomic Bloch equations \cite%
{Klim83,STH99,STH00}.

(ii) If both $\alpha $ and $\alpha ^{\prime }$ denote continuum (scattering)
states, processes like (inverse) bremsstrahlung are described.

(iii) In the case most interesting for our investigations, however, $\alpha $
denotes a bound state and $\alpha ^{\prime }$ a scattering state (or vice
versa). The transitions between them, i.e. ionization and recombination
processes, are thus included in Eq.~(\ref{BSEat}).

\subsection{Plasma Bloch equations}

\label{Bloch}

\subsubsection{Derivation of the plasma Bloch equations}

Since we are especially interested in the ionization kinetics, in the
following, we do not include (de)excitation and (de)acceleration processes,
i.e., transitions between discrete states and transitions within the
scattering continuum. Therefore, we consider a system with one single bound
state and denote
\begin{equation*}
\alpha =\left\{
\begin{array}{l@{\quad}l}
b\quad \mbox{``bound''} & \mbox{bound state} \\
f\quad \mbox{``free''} & \mbox{scattering state.}%
\end{array}
\right.
\end{equation*}
We get a system of three coupled equations \cite{KSB05a}:
\begin{eqnarray}
&&i\hbar \frac{\partial }{\partial t}F_{bb}(\mathbf{P};t)=2i\int \frac{d{\bar{%
\mathbf{p}}}}{(2\pi \hbar )^{3}}\,\mathrm{Im}\left\{ \hbar \Omega _{\mathrm{R%
}}^{bf}(\mathbf{P},{\bar{\mathbf{p}}};t)F_{fb}(\mathbf{P},{\bar{\mathbf{p}}}%
;t)\right\} 
,  \label{Fbb}
%
\\\nonumber\\
&&\left[ i\hbar \frac{\partial }{\partial t}-\left( E(\mathbf{p})-E(\mathbf{p}%
^{\prime })\right) \right] F_{ff}(\mathbf{P},\mathbf{p},\mathbf{p}^{\prime
};t)  \nonumber \\
&&-\int \frac{d{\bar{\mathbf{p}}}}{(2\pi \hbar )^{3}}\left[ \hbar \Omega _{%
\mathrm{R}}^{ff}(\mathbf{P},\mathbf{p},{\bar{\mathbf{p}}};t)F_{ff}(\mathbf{P}%
,{\bar{\mathbf{p}}},\mathbf{p}^{\prime };t)-F_{ff}(\mathbf{P},\mathbf{p},{%
\bar{\mathbf{p}}};t)\hbar {\tilde{\Omega}}_{\mathrm{R}}^{ff}(\mathbf{P},{%
\bar{\mathbf{p}}},\mathbf{p}^{\prime };t)\right]  \nonumber \\
&&=\hbar \Omega _{\mathrm{R}}^{fb}(\mathbf{P},\mathbf{p};t)F_{bf}(\mathbf{P},%
\mathbf{p}^{\prime };t)-F_{fb}(\mathbf{P},\mathbf{p};t)\hbar {\tilde{\Omega}}%
_{\mathrm{R}}^{bf}(\mathbf{P},\mathbf{p}^{\prime };t)
,  \label{Fff}
%
\\\nonumber\\
&&\left[ i\hbar \frac{\partial }{\partial t}-\left( E(\mathbf{p})-{\tilde{E}}%
_{b}\right) \right] F_{fb}(\mathbf{P},\mathbf{p};t)-\int \frac{d{\bar{%
\mathbf{p}}}}{(2\pi \hbar )^{3}}\hbar \Omega _{\mathrm{R}}^{ff}(\mathbf{P},%
\mathbf{p},{\bar{\mathbf{p}}};t)F_{fb}(\mathbf{P},{\bar{\mathbf{p}}};t)
\nonumber \\
&&=\hbar \Omega _{\mathrm{R}}^{fb}(\mathbf{P},\mathbf{p};t)F_{bb}(\mathbf{P}%
;t)-\int \frac{d{\bar{\mathbf{p}}}}{(2\pi \hbar )^{3}}F_{ff}(\mathbf{P},%
\mathbf{p},{\bar{\mathbf{p}}};t)\hbar {\tilde{\Omega}}_{\mathrm{R}}^{fb}(%
\mathbf{P},{\bar{\mathbf{p}}};t)
,\label{Ffb}
%
\\\nonumber\\
&&F_{bf}(\mathbf{P},\mathbf{p};t)=\left[ F_{fb}(\mathbf{P},\mathbf{p};t)\right]
^{\ast },  \label{Fbf}
\end{eqnarray}
where ${\tilde{E}}_{b}$ denotes the binding energy renormalized by
Hartree--Fock and Pauli blocking contributions,
\begin{eqnarray}
{\tilde{E}}_{b} &=&E_{b}+
\int \frac{d\mathbf{p}}{(2\pi \hbar )^{3}}\,|\varphi_b(\mathbf{p})|^2\,
\Big\{\Sigma_1^{\rm HF}(\mathbf{P},\mathbf{p};t)+\Sigma_2^{\rm HF}(\mathbf{P},\mathbf{p};t)
\nonumber\\
&&\hspace*{12ex}\mp \left[E(\mathbf{p})-E_b\right]\,\left[F_1(\mathbf{P},\mathbf{p};t)+F_2(\mathbf{P},\mathbf{p};t)\right]
\Big\}
%
\end{eqnarray}
with $F_{1/2}(\mathbf{P},\mathbf{p};t)=F_{1/2}\left(\frac{m_{1/2}}{M}\mathbf{P}\pm \mathbf{p};t\right)$ and $\Sigma^{\rm HF}_{1/2}(\mathbf{P},\mathbf{p};t)=
\Sigma^{\rm HF}_{1/2}\left(\frac{m_{1/2}}{M}\mathbf{P}\pm \mathbf{p};t\right)$. Here, $\varphi_b(\mathbf{p})=\langle\mathbf{p}\vert n\rangle$ is the Fourier transform of the bound state wave function.

Keeping in mind the relation (\ref{Fbf}), we have obtained a system of three equations for three unknown functions the physical meaning of which is quite obvious. $F_{bb}$ is, in principle, equivalent to the distribution function of
the atoms in the plasma. $F_{ff}$ is the binary distribution of the unbound
electron--ion pairs. From the system (\ref{Fbb}--\ref{Fbf}) we see that the
dynamics of both distributions is driven by the function $F_{fb}$ describing
the transition between bound and continuum states. The latter quantity is
closely connected with the polarization function. Its time evolution, in
turn, is determined by the distributions. Notice that, even if we neglect three-particle
collisions, ionization and recombination takes place due to
the presence of the external field contained in the Rabi energies $\hbar
\Omega _{\mathrm{R}}$.

\subsubsection{Field terms. Renormalized Rabi energies}

As we have seen before, the matrix elements of the field contribution to the
Hamiltonian $U_{ab}$ can be represented in the form of Rabi energies $\hbar
\Omega _{\mathrm{R}}$. Since we consider electron--ion pairs, $U_{ab}$ from
Eq.~(\ref{hamilt}) simplifies to
\begin{equation*}
U_{ab}=-\frac{e}{\mu}\mathbf{A}\cdot \mathbf{p}.
\end{equation*}%
Thus, the Rabi energies are given by
\begin{eqnarray}
&&\hspace*{-2ex}\hbar \Omega _{\mathrm{R}}^{ff}(\mathbf{P},\mathbf{p},\mathbf{p%
}^{\prime };t)=-\frac{e}{\mu}\mathbf{A}(t)\cdot \left\langle +\mathbf{p}%
\vert {\mathbf{p}}\vert \mathbf{p}^{\prime }+\right\rangle
+\int \frac{d\bar{\mathbf{p}}}{(2\pi \hbar)^{3}}
\varphi^*_{\mathbf{p}+}(\bar{\mathbf{p}})\varphi_{\mathbf{p}'+}(\bar{\mathbf{p}})
\nonumber\\
&&
\times\left\{
\Sigma_1^{\rm HF}(\mathbf{P},\bar{\mathbf{p}};t)+\Sigma_2^{\rm HF}(\mathbf{P},\bar{\mathbf{p}};t)
\mp \left[E(\bar{\mathbf{p}})-E({\mathbf{p}})\right]\,
\left[F_1(\mathbf{P},\bar{\mathbf{p}};t)+F_2(\mathbf{P},\bar{\mathbf{p}};t)\right]
\right\}
,  \label{Rabi1} \\%
[0.1cm]
&&\hspace*{-2ex}\hbar \Omega _{\mathrm{R}}^{fb}(\mathbf{P},\mathbf{p};t)=-%
\frac{e}{\mu}\mathbf{A}(t)\cdot \left\langle +\mathbf{p}\right\vert {%
\mathbf{p}}\left\vert n\right\rangle
+\int \frac{d\bar{\mathbf{p}}}{(2\pi \hbar)^{3}}
\varphi^*_{\mathbf{p}+}(\bar{\mathbf{p}})\varphi_{b}(\bar{\mathbf{p}})
\nonumber\\
&&
\times\left\{
\Sigma_1^{\rm HF}(\mathbf{P},\bar{\mathbf{p}};t)+\Sigma_2^{\rm HF}(\mathbf{P},\bar{\mathbf{p}};t)
\mp \left[E(\bar{\mathbf{p}})-E({\mathbf{p}})\right]\,
\left[F_1(\mathbf{P},\bar{\mathbf{p}};t)+F_2(\mathbf{P},\bar{\mathbf{p}};t)\right]
\right\}
\label{Rabi2}
\end{eqnarray}%
with $\varphi_{\mathbf{p}+}(\mathbf{p})$ being the Fourier transform of the scattering wave function.
These relations show that the bare field terms are renormalized by many-particle effects in form of
Hartree--Fock energies and Pauli blocking contributions.


\subsection{Analysis of the transition function}

\label{trans}



In order to investigate the full ionization kinetics, the system (\ref{Fbb}--%
\ref{Fbf}) has to be solved selfconsistently. This is, of course, a very
complicated task especially due to the momentum dependences of the
quantities which is, up to now, numerically not feasible. Therefore, we
approach the problem by analyzing limiting cases which simplify the system (%
\ref{Fbb}--\ref{Fbf}) and allow for an insight into the underlying physics.

Let us assume a monochromatic external field given by Eq.~(\ref{monochrom}) and look at Eq.~(\ref{Ffb}). Introducing the approximations (i) neglect of
Hartree--Fock and Pauli blocking renormalizations, (ii) replacement of
scattering states $\left|\mathbf{p}+\right>$ by free momentum states $\left|%
\mathbf{p}\right>$, and (iii) neglect of binary correlations in the
scattering state, i.e.,
\begin{equation*}
F_{ff}(\mathbf{P},\mathbf{p},\mathbf{p}^{\prime };t)\approx F_{1}(\mathbf{P%
},\mathbf{p};t)F_{2}(\mathbf{P},\mathbf{p};t)(2\pi \hbar )^{3} \delta (%
\mathbf{p}-\mathbf{p}^{\prime }),
\end{equation*}
this equation can be written as
\begin{eqnarray}  \label{Ffb1}
&&\left[i\hbar\frac{\partial}{\partial t}- \left(E(\mathbf{p})-E_{b}\right)+
\frac{e}{\mu}\mathbf{A}(t)\cdot\mathbf{p} \right] F_{fb}(\mathbf{P},%
\mathbf{p};t)  \nonumber \\
&=&-\frac{e}{\mu}\mathbf{A}(t)\cdot\mathbf{p}\,\varphi_b(\mathbf{p}) \lbrace
F_{bb}(\mathbf{P};t)-F_{1}(\mathbf{P},{\mathbf{p}};t)F_{2}(\mathbf{P},{%
\mathbf{p}};t) \rbrace.
\end{eqnarray}
The formal solution of Eq.~(\ref{Ffb1}) is given by
\begin{eqnarray}  \label{Ffb1formal}
F_{fb}(\mathbf{P},\mathbf{p};t) &=&-\frac{1}{i\hbar }\frac{e}{\mu}%
\int\limits_{t_{0}}^{t}d{\bar{t}}\, e^{-\frac{i}{\hbar }\left( E(\mathbf{p}%
)-E_{b}\right) (t-{\bar{t}})} e^{\frac{i}{\hbar}\frac{e}{\mu}\mathbf{p}%
\cdot \int\limits_{{\bar t}}^{t}d\overline{\overline{t}} \mathbf{A}(%
\overline{\overline{t}})} \mathbf{A}({\bar t})\cdot\mathbf{p}\,\varphi_b(%
\mathbf{p})  \nonumber \\
&&\times\lbrace F_{bb}(\mathbf{P};{\bar t})-F_{1}(\mathbf{P},{\mathbf{p}};{%
\bar t})F_{2}(\mathbf{P},{\mathbf{p}};{\bar t}) \rbrace.
\end{eqnarray}

For monochromatic fields (Eqs.~(\ref{monochrom})), the integral in the
second exponent on the r.h.s. of (\ref{Ffb1formal}) can be solved:
\begin{eqnarray}  \label{intAsol}
\frac{e}{\mu}\mathbf{p}\cdot \int\limits_{{\bar t}}^{t}d\overline{%
\overline{t}} \mathbf{A}(\overline{\overline{t}})= \frac{\mathbf{v}_0\cdot%
\mathbf{p}}{\omega} \left(\cos\omega t-\cos\omega{\bar t}\right).
\end{eqnarray}

Keeping in mind that the field is switched on adiabatically, i.e. at $%
t_{0}\rightarrow -\infty $, we have to write for the vector potential
\begin{equation*}
\mathbf{A}(t)=\lim\limits_{\epsilon \rightarrow 0}\mathbf{A}_{0}\frac{1}{2i}%
\left[ e^{i(\omega -i\epsilon )t}-e^{-i(\omega +i\epsilon )t}\right] .
\end{equation*}

Assuming that the time dependence of the field amplitude and the
distribution functions is weak compared to the very fast field
oscillations, they can be taken out of the integral in (\ref{Ffb1formal}).
Then, using the relation
\begin{equation*}
e^{\pm iz\cos \omega t}=\sum\limits_{n=-\infty }^{\infty }(\pm
i)^{n}J_{n}(z)e^{\pm in\omega t},
\end{equation*}%
we obtain
\begin{eqnarray}
&&F_{fb}(\mathbf{P},\mathbf{p};t)=\frac{1}{2\hbar }\frac{e}{\mu}e^{-\frac{%
i}{\hbar }\left( E(\mathbf{p})-E_{b}\right) t}\sum\limits_{k=-\infty
}^{\infty }i^{k}J_{k}\left( \frac{\mathbf{v}_{0}\cdot \mathbf{p}}{\hbar
\omega }\right) e^{ik\omega t}  \nonumber  \label{Ffb1formal1} \\
&&\times \sum\limits_{n=-\infty }^{\infty }(-i)^{n}J_{n}\left( \frac{\mathbf{%
v}_{0}\cdot \mathbf{p}}{\hbar \omega }\right) \int\limits_{t_{0}}^{t}d{\bar{t%
}}\,e^{\frac{i}{\hbar }\left( E(\mathbf{p})-E_{b}\right) {\bar{t}}%
}e^{-in\omega {\bar{t}}}\left[ e^{i(\omega -i\epsilon ){\bar{t}}%
}-e^{-i(\omega +i\epsilon ){\bar{t}}}\right]   \nonumber \\
&&\times \mathbf{A}_{0}\cdot \mathbf{p}\,\varphi_b(\mathbf{p})\left[ F_{bb}(%
\mathbf{P};t)-F_{1}(\mathbf{P},{\mathbf{p}};t)F_{2}(\mathbf{P},{\mathbf{p%
}};t)\right] .
\end{eqnarray}%
The time integration can be carried out, and after an index shift $k=n+l$ we
obtain the result
\begin{eqnarray}
&&F_{fb}(\mathbf{P},\mathbf{p};t)=-\frac{e}{2\mu}\sum\limits_{n=-\infty
}^{\infty }\sum\limits_{l=-\infty }^{\infty }i^{l-1}J_{n}\left( \frac{\mathbf{v%
}_{0}\cdot \mathbf{p}}{\hbar \omega }\right) J_{n+l}\left( \frac{\mathbf{v}%
_{0}\cdot \mathbf{p}}{\hbar \omega }\right)   \nonumber  \label{Ffb1sol} \\
&&\times \left[ \frac{e^{i((l+1)\omega -i\epsilon )t}}{E_{b}-E(\mathbf{p}%
)+(n-1)\hbar \omega +i\hbar \epsilon }-\frac{e^{i((l-1)\omega -i\epsilon )t}%
}{E_{b}-E(\mathbf{p})+(n+1)\hbar \omega +i\hbar \epsilon }\right]   \nonumber \\
&&\times \mathbf{A}_{0}\cdot \mathbf{p}\,\varphi_b(\mathbf{p})\left[ F_{bb}(%
\mathbf{P};t)-F_{1}(\mathbf{P},{\mathbf{p}};t)F_{2}(\mathbf{P},{\mathbf{p%
}};t)\right] .
\end{eqnarray}%
This formula can be rewritten once more by appropriate index shifts and
using the Bessel function relations
\begin{equation}
J_{-n}(z)=(-1)^{n}J_{n}(z)\ ;\ \ \ \ J_{n-1}(z)+J_{n+1}(z)=\frac{2n}{z}%
J_{n}(z).  \label{Bess}
\end{equation}%
We arrive at
\begin{eqnarray}\label{Ffbres}
F_{fb}(\mathbf{P},\mathbf{p};t)=\frac{e}{\mu}\sum\limits_{n=-\infty
}^{\infty }\sum\limits_{l=-\infty }^{\infty }i^{l}\frac{n\hbar \omega }{%
\mathbf{v}_{0}\cdot \mathbf{p}}J_{n}\left( \frac{\mathbf{v}_{0}\cdot \mathbf{%
p}}{\hbar \omega }\right) J_{n+l}\left( \frac{\mathbf{v}_{0}\cdot \mathbf{p}%
}{\hbar \omega }\right)   \nonumber \\
\times\frac{e^{i(l\omega -i\epsilon )t}}{E_{b}-E(\mathbf{p})+n\hbar \omega
+i\hbar \epsilon }\mathbf{A}_{0}\cdot \mathbf{p}\,\varphi_b(\mathbf{p})\left[
F_{bb}(\mathbf{P};t)-F_{1}(\mathbf{P},{\mathbf{p}};t)F_{2}(\mathbf{P},{%
\mathbf{p}};t)\right] .
\end{eqnarray}

As can be seen from this result, the field causes interesting physical
effects. First, the sum over $l$ indicates the generation of higher
harmonics of the field frequency, cf. the term $e^{i(l\omega-i\epsilon)t}$.
On the other hand, the sum over $n$ reflects the absorption or emission of
multiple photons, i.e. multiphoton ionization, which finds its expression in
the denominator $E_b-E(\mathbf{p})+n\hbar\omega+i\hbar\epsilon$. Similar
effects have been found, e.g., in \cite{kremp99,bonitz99,BSHK01}.

\subsection{Absorption and ionization}\label{absion}

\subsubsection{Electrical current and absorption coefficient}

Now we are able to consider the current density and other interesting
physical quantities connected to $\mathbf{j}$ like polarization,
absorption and ionization coefficients. For this purpose, let us come back to
the balance equation (\ref{Currbal}).

With the completeness relation for the eigenstates $\left\vert
\mathbf{P}\alpha \right\rangle$ and the Schr\"{o}dinger equation
\begin{equation*}
1=\sum\limits_{\mathbf{P}\alpha }\left\vert \mathbf{P}\alpha \right\rangle
\left\langle \mathbf{P}\alpha \right\vert, \ \ \ \ \ \ \ \ \ \ \left(
H_{ab}^{0}-E_{\alpha \mathbf{P}}\right) \left\vert \mathbf{P}\alpha
\right\rangle =-V_{ab}\left\vert \mathbf{P}\alpha \right\rangle ,
\end{equation*}
we express the right hand side (first line) of the balance equation (\ref{Currbal}) for the
current density in terms of the $F_{ab}^{\alpha \alpha ^{\prime }}(\mathbf{P};t)$. It follows
\begin{eqnarray}
\frac{d\mathbf{j}(t)}{dt}-\omega _{pl}^{2}\mathbf{E}=-\frac{\mathcal{V}}{2i\hbar}%
\sum\limits_{ab}\int \frac{d\mathbf{p}}{(2\pi \hbar )^{3}}\int \frac{d\mathbf{P}}{(2\pi \hbar )^{3}}%
\sum\limits_{\alpha \bar{\alpha }}\left(\frac{e_{t}\mathbf{P}}{M}+\frac{e_{r}%
\mathbf{p}}{\mu}\right)\ (E_{\mathbf{pP}}-E_{\alpha \mathbf{P}})  \nonumber \\
\left[ \varphi _{\alpha }(\mathbf{p})\varphi _{\bar{\alpha} }^{\ast }(%
\mathbf{p})F_{ab}^{\alpha \bar{\alpha }}(\mathbf{P};t)-\varphi _{\alpha
}^{\ast }(\mathbf{p})\varphi _{\bar{\alpha }}(\mathbf{p})F_{ab}^{\bar{\alpha}\alpha
}(\mathbf{P};t)\right]=\sum\limits_{ab}I_{ab}(t)\,.
\label{curr1}
\end{eqnarray}

This equation shows that the different transition processes discussed
in Sec.~\ref{basic} contribute to the current in a partially ionized plasma. In the following, we consider a hydrogen plasma as a simple model case. We should notice, however, that the theory can be applied to more complicated systems, too. We consider here
the electron--proton part and especially the bound--free contribution ($\alpha
=n=1,\bar{\alpha }=\mathbf{p}$ and vice versa)
\begin{equation}
I_{ep}(t)=\frac{\mathcal{V}}{\hbar}
\int \frac{d\mathbf{p}}{(2\pi \hbar )^{3}}\int \frac{d\mathbf{P}}{(2\pi \hbar )^{3}}
\frac{e\mathbf{p}}{\mu}\,(E(\mathbf{p})-E_b)\,\varphi_b(\mathbf{p})\,{\rm Im}\,F_{fb}(\mathbf{P};t).
\label{curr2}
\end{equation}

Then we introduce the expression (\ref{Ffbres}) into (\ref{curr2}) and ignore the weak
time dependence of the distribution functions and the field amplitude.
Finally we integrate Eq.~(\ref{curr1}) over time and obtain
\begin{equation}
\mathbf{j}(t)=\mathbf{j}_{0}(t)+\sum\limits_{l=-\infty
}^{\infty }\mathbf{j}_{l}(\omega)e^{-il\omega t}
\label{jfour1}
\end{equation}
which is clearly the Fourier expansion of the current in terms of all
harmonics of the field frequency $\omega$. The Fourier coefficients
$\mathbf{j}_{l}(\omega)$ are given by
\begin{eqnarray}
\mathbf{j}_{l}(\omega)&=&-\frac{\mathcal{V}}{2}
\int \frac{d\mathbf{p}}{(2\pi \hbar )^{3}}\int \frac{d\mathbf{P}}{(2\pi \hbar )^{3}}
\frac{e}{\mu }\,\frac{E(\mathbf{p})-E_b}{l\hbar \omega }
\,\mathbf{p}\,\varphi_b(\mathbf{p})\,
\frac{e}{\mu}\mathbf{A}_0\cdot \mathbf{p}\,\varphi_b(\mathbf{p})  \nonumber
\\
&&\times\left[ F_{bb}(\mathbf{P};t)-F_{e}(\mathbf{P},{\mathbf{p}};t)F_{p}(%
\mathbf{P},{\mathbf{p}};t)\right]\sum\limits_{n=-\infty }^{\infty }i^{l}\frac{%
n\hbar \omega }{\mathbf{v}_{0}\cdot \mathbf{p}}J_{n}\left( \frac{\mathbf{v}%
_{0}\cdot \mathbf{p}}{\hbar \omega }\right) J_{n-l}\left( \frac{\mathbf{v}%
_{0}\cdot \mathbf{p}}{\hbar \omega }\right)  \nonumber \\
&&\times\left[ \frac{(-1)^{l}}{E_{b}-E(\mathbf{p})+n\hbar \omega +i\hbar \epsilon }+%
\frac{1}{E_{b}-E(\mathbf{p})-n\hbar \omega -i\hbar \epsilon }\right]  \label{jl}
\end{eqnarray}
Furthermore, $\mathbf{j}_{0}(t)=\omega _{pl}^{2}$\ $\int\limits_{-\infty}^{t} dt'
\mathbf{E}(t')$ is the current of the collisionless plasma.

As already mentioned above, the current determines several physical quantities, e.g. the polarization. Here, we will consider the energy transfer $\mathbf{j}\cdot \mathbf{E}$ between the field and the plasma. For the field $\mathbf{E}$ we assume the time
dependence (\ref{monochrom}). We consider the dissipation of the energy
averaged over one cycle of oscillation. Then we obtain after a simple
calculation
\begin{equation}
\left\langle \mathbf{j}\cdot \mathbf{E}\right\rangle =\frac{1}{T}%
\int\limits_{t-T}^{t}dt^{\prime }\,\mathbf{j}(t)\cdot \mathbf{E}(t)=\mathbf{E}_{0}\cdot {\rm Re}\, \mathbf{j}_{1}(\omega).
\label{jav}
\end{equation}

Now we introduce the absorption coefficient for the bound--free transition by
\begin{equation}
\alpha _{bf}(\omega )=\frac{1}{c\,\varepsilon _{0}}\frac{\left\langle
\mathbf{j}\cdot \mathbf{E}\right\rangle }{\left\langle \mathbf{E}%
^{2}\right\rangle }  \label{abs}
\end{equation}

With the help of the Dirac identity, the substitution $\ n\rightarrow -n$ in
the second term and the Bessel function relations (\ref{Bess}),
after some algebra follows
\begin{eqnarray}
&&\alpha _{bf}(\omega )=\frac{\mathcal{V}}{c\,\varepsilon _{0}\hbar }\frac{%
2\pi }{E_{0}^{2}}\sum\limits_{n=-\infty }^{\infty }(n\hbar \omega )^{3}
\int \frac{d\mathbf{p}}{(2\pi \hbar )^{3}}\int \frac{d\mathbf{P}}{(2\pi \hbar )^{3}}\frac{E({\mathbf{p})%
}-E_{b}}{n\hbar \omega }\left\vert \varphi_b(\mathbf{p})\right\vert^{2}  \nonumber \\
&&\times J_{n}^{2}\left( \frac{\mathbf{v}_{0}\cdot \mathbf{%
p}}{\hbar \omega }\right) \delta (E(\mathbf{p})-E_{b}-n\hbar \omega)
\left[F_{bb}(\mathbf{P};t)-F_{e}(\mathbf{P},{\mathbf{p}};t)F_{p}(%
\mathbf{P},{\mathbf{p}};t)\right]
\label{abs1}
\end{eqnarray}
The quantity $\alpha_{bf}(\omega )$ describes the process of the absorption (emission) of $n$
photons in the ionization (recombination) of atoms. In order to evaluate Eq.~(\ref{abs1}), the knowledge of the time evolution of the distribution functions of atoms and free particles is necessary. This evolution has, in principle, to be described by the equations (\ref{Fbb}) and (\ref{Fff}). In order to obtain first results, we replace these quantities by thermodynamic equilibrium functions, i.e., Maxwellian distributions. This assumption represents a significant simplification, in particular, the influence of the field on the bound states is neglected. An improved determination of $F_{bb}$ from Eq.~(\ref{Fbb}) leads to a Stark shift of levels quadratic in the $\mathbf{A}\cdot\mathbf{p}$ interaction as was discussed also by Milonni and Ackerhalt \cite{MA89}.

\begin{figure}[bth]
\centerline{\epsfig{figure=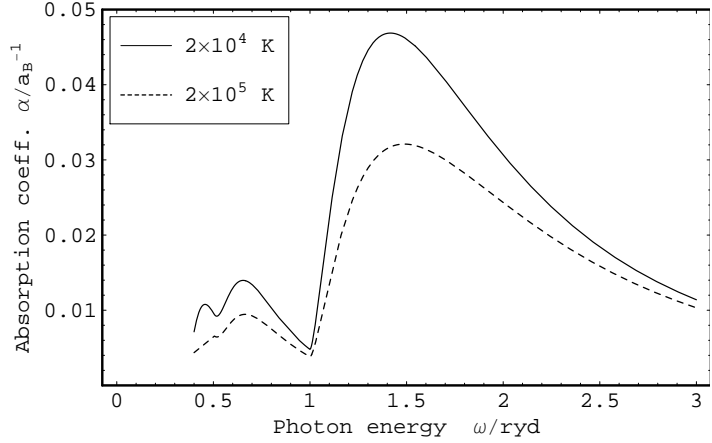,clip=true,width=10cm}}
\caption{Absorption coefficient $\alpha_{bf}$ vs. laser frequency for two different temperatures.\label{abs12}}
\end{figure}

The absorption coefficient $\alpha_{bf}(\omega )$ is shown in Fig.~\ref{abs12} in dependence on the laser frequency. An interesting feature is the occurence of additional maxima for photon energies $\hbar\omega$ smaller than the ionization energy. They are caused by two- and more-photon processes.

If we restrict the sum in formula (\ref{abs1}) to $n=1$, our approach reproduces the relations of linear response theory. 

\subsubsection{Rate coefficients}

The absorption of photons is, under the condition $\ E(\mathbf{p}%
)-E_{b}=n\hbar \omega$, connected with a bound--free transition, i.e. the
ionization of the atoms. The process of ionization, of course, changes the
plasma composition. From the macroscopic point of view, the time dependence
of the densities of the plasma particles is given by the rate equation (\ref{rate}). If we restrict ourselves to radiative processes, the change of the atomic density $n_{A}(t)$ is given by (in the following, the superscript $R$ will be dropped)
\begin{equation}
\frac{\partial }{\partial t}n_{A}(t)=-\alpha (t)n_{A}(t)+\beta
(t)n_{e}(t)n_{p}(t).  \label{reqn}
\end{equation}%
Here, $n_{e}(t)$ and $n_{p}(t)$ are the densities of free particles (electrons and protons), and the coefficients $\alpha (t)$ and $\beta (t)$ are the radiative rate coefficients.

From the microscopic point of view, the rate equation (\ref{reqn}) follows
from the equation for the occupation of the atoms $F_{bb}$ (\ref{Fbb}),
\begin{eqnarray}
\frac{\partial }{\partial t}F_{bb}(\mathbf{P};t)=\frac{2}{\hbar }\mathrm{sin}\,\omega t\int \frac{d{\mathbf{p}}}{(2\pi \hbar )^{3}}{\bf v}_0\cdot{\bf p}\,\varphi_b({\bf p})\,
\mathrm{Im}\,F_{fb}(\mathbf{P},{\bf p};t).
\end{eqnarray}
Inserting the solution (\ref{Ffbres}) and using the Dirac identity, we arrive at
\begin{eqnarray}\label{Fbbres2}
&&\frac{\partial }{\partial t}F_{bb}(\mathbf{P};t)=-\frac{2}{\hbar }\mathrm{sin}\,\omega t\int \frac{d{\mathbf{p}}}{(2\pi \hbar )^{3}}{\bf v}_0\cdot{\bf p}\,\left|\varphi_b({\bf p})\right|^2
\sum\limits_{n=-\infty}^{\infty}\sum\limits_{l=-\infty}^{\infty}
n\hbar\omega\,J_n\left(\frac{\mathbf{v}_{0}\cdot \mathbf{p}}{\hbar \omega }\right)
J_{n+l}\left( \frac{\mathbf{v}_{0}\cdot \mathbf{p}}{\hbar \omega }\right)\nonumber\\
&&\times
\Bigg\{\sin\left[l\left(\omega t+\frac{\pi}{2}\right)\right]\frac{P}{E_{b}-E(\mathbf{p})+n\hbar \omega }
-\pi \cos\left[l\left(\omega t+\frac{\pi}{2}\right)\right] \delta(E(\mathbf{p})-E_{b}-n\hbar \omega ) \Bigg\}  \nonumber \\
&&\times\left[ F_{bb}(\mathbf{P};t)-F_{e}(\mathbf{P}%
,{\mathbf{p}};t)F_{p}(\mathbf{P},{\mathbf{p}};t)\right].
\end{eqnarray}

In the following, we assume Maxwellian distributions for the free particles,
i.e.
\begin{equation*}
F_{e/p}(\mathbf{P},\mathbf{p};t)=\frac{n_{e/p}(t)\Lambda _{e/p}^{3}}{2}%
\,e^{-\frac{1}{2m_{e/p}k_{\mathrm{B}}T}\left( \frac{m_{e/p}}{M}\mathbf{P}\pm
\mathbf{p}\right) ^{2}},
\end{equation*}%
with the thermal wavelength $\Lambda _{e/p}=\sqrt{\frac{2\pi \hbar ^{2}}{%
m_{e/p}k_{\mathrm{B}}T}}$. Integrating Eq.~(\ref{Fbbres2}) over $\mathbf{P}$, we obtain the rate equation (\ref{reqn}) for the density of the atoms $n_{A}(t)$. Obviously, the ionization coefficient $\alpha (t)$ is then given by the microscopic expression
\begin{eqnarray}
&&\alpha (t)=
\frac{2}{\hbar }\,\mathrm{sin}\,\omega t\int \frac{d{\mathbf{p}}}{(2\pi \hbar )^{3}}{\bf v}_0\cdot{\bf p}\,\left|\varphi_b({\bf p})\right|^2
\sum\limits_{n=-\infty}^{\infty}\sum\limits_{l=-\infty}^{\infty}
n\hbar\omega\,J_n\left(\frac{\mathbf{v}_{0}\cdot \mathbf{p}}{\hbar \omega }\right)
J_{n+l}\left( \frac{\mathbf{v}_{0}\cdot \mathbf{p}}{\hbar \omega }\right)\nonumber\\
&&\times
\Bigg\{\sin\left[l\left(\omega t+\frac{\pi}{2}\right)\right]\frac{P}{E_{b}-E(\mathbf{p})+n\hbar \omega }
-\pi \cos\left[l\left(\omega t+\frac{\pi}{2}\right)\right] \delta(E(\mathbf{p})-E_{b}-n\hbar \omega ) \Bigg\},
\label{ionc}
\end{eqnarray}
and the recombination coefficient follows from
\begin{eqnarray}
&&\beta (t)=
\frac{\Lambda _{r}^{3}}{2\hbar }\,\mathrm{sin}\,\omega t\int \frac{d{\mathbf{p}}}{(2\pi \hbar )^{3}}{\bf v}_0\cdot{\bf p}\,\left|\varphi_b({\bf p})\right|^2
\sum\limits_{n=-\infty}^{\infty}\sum\limits_{l=-\infty}^{\infty}
n\hbar\omega\,J_n\left(\frac{\mathbf{v}_{0}\cdot \mathbf{p}}{\hbar \omega }\right)
J_{n+l}\left( \frac{\mathbf{v}_{0}\cdot \mathbf{p}}{\hbar \omega }\right)\nonumber\\
&&\times
\Bigg\{\sin\left[l\left(\omega t+\frac{\pi}{2}\right)\right]\frac{P}{E_{b}-E(\mathbf{p})+n\hbar \omega }
-\pi \cos\left[l\left(\omega t+\frac{\pi}{2}\right)\right] \delta(E(\mathbf{p})-E_{b}-n\hbar \omega ) \Bigg\} 
\nonumber \\
&&\hspace{3ex}\times
e^{-\frac{p^{2}}{2\mu k_{\mathrm{B}}T}}
\end{eqnarray}
with $\Lambda _{r}=\sqrt{\frac{2\pi \hbar ^{2}}{\mu k_{\mathrm{B}}T}}$. It
is useful to average again over one period of the field oscillation. Using the relations
$\frac{1}{T}\int\limits_0^Tdt\,\sin\omega t\cos l\omega t=0$,
$\frac{1}{T}\int\limits_0^Tdt\,\sin\omega t\sin l\omega t=\frac{l}{2}\delta_{|l|1}$,
we get the averaged coefficients
\begin{eqnarray}
&&\bar\alpha=
\frac{2\pi}{\hbar }\int \frac{d{\mathbf{p}}}{(2\pi \hbar )^{3}}\,\left|\varphi_b({\bf p})\right|^2\sum\limits_{n=-\infty}^{\infty}
(n\hbar\omega)^2\,J_n^2\left(\frac{\mathbf{v}_{0}\cdot \mathbf{p}}{\hbar \omega }\right)
\delta(E(\mathbf{p})-E_{b}-n\hbar \omega ),
\label{ionc1}\\
&&\bar\beta=
\frac{\pi\Lambda _{r}^{3}}{2\hbar }\int \frac{d{\mathbf{p}}}{(2\pi \hbar )^{3}}\,\left|\varphi_b({\bf p})\right|^2\sum\limits_{n=-\infty}^{\infty}
(n\hbar\omega)^2\,J_n^2\left(\frac{\mathbf{v}_{0}\cdot \mathbf{p}}{\hbar \omega }\right)
\delta(E(\mathbf{p})-E_{b}-n\hbar \omega ) e^{-\frac{p^{2}}{2\mu k_{\mathrm{B}}T}}.
\end{eqnarray}

\begin{figure}[bth]
\centerline{\epsfig{figure=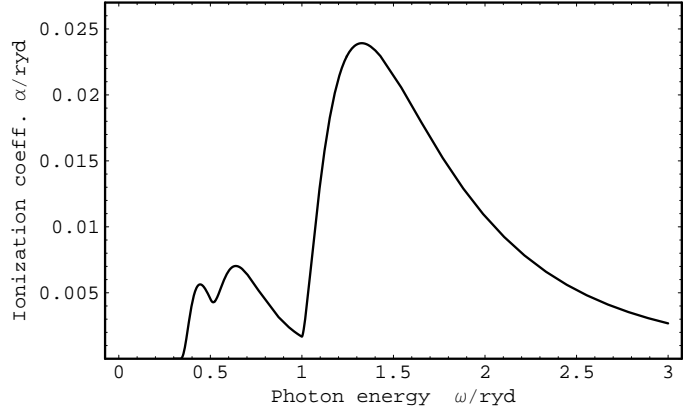,clip=true,width=9cm}}
\caption{Ionization coefficient ${\bar\alpha}$ vs. laser frequency.\label{alpha12}}
\end{figure}
This strongly simplified expression for the ionization coefficient is essentially equivalent to results from atomic physics \cite{keldysh64,faisal73,reiss80,MA89}. However, in contrast to the results of Faisal and Reiss, no effect of the $A^2$ term occurs. Effects of higher order in $A$ following from the ${\bf p}\cdot{\bf A}$ coupling are contained, in principle, in the basic equations (\ref{Fbb}--\ref{Fbf}) as mentioned already above in the discussion of the absorption coefficient.

In Fig.~\ref{alpha12} we show the ionization coefficient ${\bar\alpha}$ as a function of the photon energy. Similar to the absorption coefficient we again observe a finite ionization probability for photon energies smaller than the binding energy due to multi-photon processes.
However, we want to remark that this multi-photon interpretation of the ionization process is limited by the Keldysh frequency \cite{keldysh64}
\begin{eqnarray}
\omega _{K}=\frac{eE}{\sqrt{2mI}},
\end{eqnarray}
where $I$ is the ionization energy. Only for frequencies larger than $\omega _{K}$ the multi-photon interpretation is meaningful. For frequencies smaller than $\omega _{K}$ the ionization has to be interpreted by the usual tunnel effect.

\ack

This work was supported by the Deutsche Forschungsgemeinschaft (SFB 198).


\section*{References}

\begin{thebibliography}{99}
\bibitem{perry94} M.D. Perry and G. Mourou, Science \textbf{264}, 917 (1994)

\bibitem{lindl04} J.D. Lindl, P. Amendt, R.L. Berger \textit{et al.}, Phys. Plasmas {\bf 11}, 340 (2004)

\bibitem{KSK2005}D. Kremp, M. Schlanges, and W.-D. Kraeft, \textit{Quantum Statistics of
Nonideal Plasmas} (Springer, Berlin 2005)

\bibitem{SBK88} M. Schlanges, Th. Bornath, and D. Kremp, Phys. Rev. A \textbf{38}, 2174 (1988)

\bibitem{BS93} Th. Bornath and M. Schlanges, Physica A \textbf{196}, 427 (1993)

\bibitem{kremp99} D. Kremp, Th. Bornath, M. Bonitz, and M. Schlanges, Phys. Rev. E
\textbf{60}, 4725 (1999)

\bibitem{bonitz} M. Bonitz, {\em Quantum Kinetic Theory} (Teubner, Stuttgart/Leipzig 1998)

\bibitem{bornath03}
Th. Bornath, M. Schlanges, P. Hilse, and D. Kremp, J. Phys. A: Math. Gen. {\bf 36} 5941-5948
(2003).

\bibitem{springer_morawetz}
Th. Bornath, M. Schlanges, P. Hilse, and D. Kremp, in: K. Morawetz (Ed.)
Nonequilibrium physics at short time scales: formation of correlations (Springer-Verlag, Berlin 2004),
pp. 153-172.

\bibitem{silin64} V.P. Silin, Zh. Eksp. Teor. Fiz. \textbf{47}, 2254 (1964),
[JETP \textbf{20}, 1510 (1965)]

\bibitem{klimontovich75} Yu.L. Klimontovich, \textit{Kinetic Theory of
Nonideal Gases and Nonideal Plasmas} (russ.), (Nauka, Moscow 1975) [Engl.
transl.: Pergamon Press, Oxford 1982]

\bibitem{bonitz99} M. Bonitz, Th. Bornath, D. Kremp, M. Schlanges, and W.D.
Kraeft, Contrib. Plasma Phys. \textbf{39}, 329 (1999)

\bibitem{SKB99} D. Semkat, D. Kremp, and M. Bonitz, Phys. Rev. E \textbf{59}, 1557 (1999)

\bibitem{SKB02} D. Semkat, D. Kremp, and M. Bonitz,
Contrib. Plasma Phys. {\bf 42}, 31 (2002)

\bibitem{SBK03} D. Semkat, M. Bonitz, and D. Kremp,
Contrib. Plasma Phys. {\bf 43}, 321 (2003)

\bibitem{KSB05}
D. Kremp, D. Semkat, and M. Bonitz, J. Phys.: Conference Series \textbf{11}, 1 (2005)

\bibitem{BKS99} Th. Bornath, D. Kremp, and M. Schlanges, Phys. Rev. E
\textbf{60}, 6382 (1999)

\bibitem{oberman62} C. Oberman, A. Ron, and J. Dawson, Phys. Fluids \textbf{5%
}, 1514 (1962); J.M. Dawson and C. Oberman: Phys. Fluids \textbf{6}, 394 (1963)

\bibitem{decker94} C.D. Decker, W.B. Mori, J.M. Dawson, and T. Katsouleas,
Phys. Plasmas \textbf{1}, 4043 (1994)

\bibitem{mulser}
P. Mulser, F. Cornolti, E. B\`{e}suelle, and R. Schneider, Phys. Rev. E \textbf{63}, 63 (2000)

\bibitem{kull01} H.-J. Kull and L. Plagne, Phys. of Plasmas \textbf{8}, 5244
(2001)

\bibitem{bshk01} Th. Bornath, M. Schlanges, P. Hilse, and D. Kremp, Phys.
Rev. E \textbf{64}, 26414 (2001)

\bibitem{hazak} G. Hazak, N. Metzler, M. Klapisch, and J. Gardner, Phys.
of Plasmas \textbf{9}, 345 (2002)

\bibitem{reinholz00} H. Reinholz, R. Redmer, G. R\"opke, and A. Wierling,
Phys. Rev. E \textbf{62}, 5648 (2000)

\bibitem{bornath04}
Th. Bornath, D. Kremp, and M. Schlanges, unpublished (2004)

\bibitem{bekefi66} G. Bekefi, {\em Radiation processses in plasmas} (Wiley, New York 1966)

\bibitem{dresden}
M. Schlanges, P. Hilse, Th. Bornath, and D. Kremp, in: M. Bonitz and D. Semkat (Eds.)
Progress in Nonequilibrium Green's Functions (World Scientific, Singapore 2003), pp. 50-65.

\bibitem{haberland} H. Haberland, M. Bonitz, and D. Kremp, Phys. Rev. E \textbf{64}, 026405 (2001)

\bibitem{SBKH03} M. Schlanges, Th. Bornath, D. Kremp, and P. Hilse,
Contrib. Plasma Phys. \textbf{43}, 360 (2003)

\bibitem{bornath05}
Th. Bornath, D. Kremp, P. Hilse, and M. Schlanges, J. Phys.: Conference Series \textbf{11}, 180 (2005)

\bibitem{cauble85} R. Cauble and W. Rozmus, Phys. Fluids \textbf{28}, 3387
(1985)

\bibitem{pfalzner98} S. Pfalzner and P. Gibbon, Phys. Rev. E \textbf{57},
4698 (1998)

\bibitem{BSKF03} M. Bonitz, D. Semkat, D. Kremp, and V. S. Filinov,
in {\it Progress in Nonequilibrium Green's functions II},
M.~Bonitz and D.~Semkat (Eds.),
World Scientific Publ., Singapore 2003, pp.~445-463

\bibitem{KSB05a}
D. Semkat, D. Kremp, and M. Bonitz, J. Phys.: Conference Series \textbf{11}, 25 (2005)

\bibitem{Klim83} Yu.~L.~Klimontovich, \textit{Kinetic theory of
electromagnetic processes}, Springer, Berlin, Heidelberg, New York (1983)

\bibitem{STH99} R.~Schepe, D.~Tamme, and K.~Henneberger, Contrib. Plasma
Phys. \textbf{39}, 29 (1999)

\bibitem{STH00} Schepe R, Tamme D and Henneberger K 2000 in \textit{Progress
in Nonequilibrium Green's Functions} ed M Bonitz (Singapore: World Scientific Publ.) p 170

\bibitem{BSHK01} Bornath Th, Schlanges M, Hilse P and Kremp D 2001 \textit{Phys. Rev.} E \textbf{64} 26414 

\bibitem{keldysh64} L.~V.~Keldysh, Zh. Eksp. Teor. Fiz. \textbf{47}, 1945
(1964) [{Sov. Phys.--JETP} \textbf{20}, 1307 (1965)]

\bibitem{faisal73} F.~Faisal, J. Phys. B: Atom. Molec. Phys. \textbf{6}, L
89 (1973)

\bibitem{reiss80} H.~R.~Reiss, Phys. Rev. A \textbf{22}, 1786 (1980); Phys.
Rev. A \textbf{42}, 1476 (1989)

\bibitem{MA89} P.~W.~Milonni and J.~R.~Ackerhalt, Phys. Rev. A \textbf{39},
1139 (1989)


\end{thebibliography}
\end{document}